\documentclass[12pt]{article}

\usepackage{amssymb,amsthm,amscd,array}

\let\ssection=\section
\renewcommand{\section}{\setcounter{equation}{0}\ssection}

\newcommand{\bR}{{\bf R}}

\newcommand{\bbC}{\mathbb{C}}

\newcommand{\bone}{{\bf 1}}

\newcommand{\const}{\mathrm{const}}

\newcommand{\cH}{\mathcal{H}}
\newcommand{\cM}{{\mathcal{M}}}

\newcommand{\cE}{{\mathcal{E}}}
\newcommand{\Diff}{\mathrm{Diff}}
\newcommand{\rdiv}{\mathrm{div}}

\newcommand{\cF}{{\mathcal{F}}}

\newcommand{\GL}{{\mathrm{GL}}}

\newcommand{\rg}{\mathrm{g}}

\newcommand{\cQ}{{\mathcal{Q}}}

\newcommand{\Vect}{\mathrm{Vect}}
\newcommand{\vol}{\mathrm{vol}}

\newcommand{\cY}{\mathcal{Y}}

\newcommand{\half}{{\scriptstyle\frac{1}{2}}}
\newcommand{\fg}{\mathfrak{g}}

\newcommand{\scirc}{{\scriptstyle\circ}}
\newcommand{\la}{{\langle}}
\newcommand{\ra}{{\rangle}}

\begin{document}

\baselineskip=17pt

\def\a{\alpha}
\def\b{\beta}
\def\c{\gamma}
\def\d{\delta}
\def\g{\gamma}
\def\om{\omega}
\def\r{\rho}
\def\s{\sigma}
\def\vfi{\varphi}
\def\l{\lambda}
\def\m{\mu}
\def\implies{\Rightarrow}

\oddsidemargin .1truein
\newtheorem{thm}{Theorem}[section]
\newtheorem{lem}[thm]{Lemma}
\newtheorem{cor}[thm]{Corollary}
\newtheorem{pro}[thm]{Proposition}
\newtheorem{ex}[thm]{Example}
\newtheorem{rmk}[thm]{Remark}
\newtheorem{defi}[thm]{Definition}

\title{Quantum integrability of quadratic Killing tensors}

\author{C. DUVAL\footnote{mailto:duval@cpt.univ-mrs.fr}\\
Centre de Physique Th\'eorique, CNRS, 
Luminy, Case 907\\ 
F-13288 Marseille Cedex 9 (France)\footnote{ 
UMR 6207 du CNRS associ\'ee aux 
Universit\'es d'Aix-Marseille I et II 
et Universit\'e du Sud Toulon-Var; 
Laboratoire 
affili\'e \`a la FRUMAM-FR2291}
\and
G. VALENT\footnote{mailto:gvalent@lumimath.univ-mrs.fr}\\
Laboratoire de Physique Th\'eorique et des Hautes Energies,\\
2, Place Jussieu\\
F-75251 Paris Cedex 5 (France)\footnote{
UMR 7589 du CNRS associ\'ee aux 
Universit\'es Paris VI et Paris VII}
}

\date{}

\maketitle

\thispagestyle{empty}

\begin{abstract}
Quantum integrability of classical integrable systems given by quadratic Killing tensors on curved configuration spaces is investigated. It is proven that, using a ``minimal'' quantization scheme, quantum integrability is insured for a large class of classic examples.
\end{abstract}

\vskip4cm
\noindent
Preprint: CPT-2004/P.120 and LPTHE-04-33


\newpage


\section{Introduction}\label{Intro}

One of the main goals of this article is to present a somewhat general framework for 
the quantization of classical observables on a cotangent bundle which are polynomials 
at most cubic in momenta. This approach enables us to investigate the quantization 
of classically Poisson-commuting observables, and hence to tackle the problem of 
quantum integrability for a reasonably large class of dynamical systems.

What should actually be the definition of quantum integrability is a long standing 
issue, see, e.g., \cite{Wei}. The point of view  espoused in this paper is the following. Start with a complete set of independent Poisson-commuting classical observables, and use some quantization rule to get a corresponding set of quantum observables; if these operators appear to be still in involution with respect to the commutator, the system will be called integrable at the quantum level.

Our work can be considered as a sequel to earlier and pioneering contributions \cite{BCar3, BCR1, BCR2, HW, Tot} that provide worked examples of persistence of integrability 
from the classical to the quantum regime. The general approach we deal with in 
this paper helps us to highlight the general structure of quantum corrections and to show 
that the latter actually vanish in most, yet not all, interesting examples.

Returning to the general issue of quantization, let us mention that our choice 
of quantization procedure, which we might call ``minimal'', doesn't stem from 
first principles, e.g., from invariance or equivariance requirements involving 
some specific symmetry. Although this ``minimal'' quantization only applies to 
low degree polynomials on cotangent bundles, it has the virtue of leading 
automatically to the simplest symmetric operators that guarantee quantum 
integrability in many cases. In order to provide the explicit form of the 
quantization scheme, hence of the quantum corrections, we need a symmetric linear 
connection be given on the base of our cotangent bundle. In most examples 
where a (pseudo-)Riemannian metric is considered from the outset, this connection 
will be chosen as the Levi-Civita connection.

To exemplify our construction, we consider a number of examples of classical 
integrable systems together with their quantization. For instance, our approach for dealing with quantum integrability in somewhat general terms allowed us to deduce the quantum integrability of the Hamiltonian flow for the generalized Kerr-Newman solution of the Einstein-Maxwell equations with a cosmological constant first discovered by Carter \cite{BCar1,BCar2,BCar3}.
Also does our quantization scheme leads us to an independent proof of the quantum integrability for St\"ackel systems originally due to Benenti, Chanu and Rastelli \cite{BCR1, BCR2}. 

The paper is organised as follows. In Section \ref{ClassIntSys} we gather the definitions of the Schouten bracket of symmetric 
contravariant tensor fields on configuration space, $M$.  We make use of Souriau's 
procedure to present, in a manifestly gauge invariant fashion, the minimal coupling 
to an external electromagnetic field; this enables us to provide a geometric 
definition of the so called Schouten-Maxwell bracket. The related definitions 
of Killing and Killing-Maxwell tensors follow naturally and will be used throughout 
the rest of the paper.
We recall the basics of classical integrable systems, with emphasis on the St\"ackel 
class. The main objective of the present Section is then to revisit some classic examples of integrable systems involving Killing tensors. Naturally starting with the Jacobi system on the ellipsoid, we prove, {\it en passant}, that it is locally of the St\"ackel type, even allowing for an extra harmonic potential. This extends previous work of Benenti \cite{Ben} related to the geodesic flow of the ellipsoid. Similarly, we show that the Neumann system is also locally St\"ackel. A number of additional examples, not of St\"ackel type, e.g., the Di Pirro system, and the geodesic flow on various (pseudo-)Riemannian manifolds such as the Kerr-Newman-de Sitter solution and the Multi-Centre solution are also considered.

We introduce, in Section \ref{QuantIntSys}, a specific ``minimal'' quantization scheme 
for observables at most cubic in momenta on the cotangent bundle $T^*M$ of a smooth manifold $M$  endowed with a symmetric connection $\nabla$, extending a 
previous proposal \cite{BCar3}. This quantization mapping is shown to be equivariant with respect to the affine group of $(M,\nabla)$. The computation of the commutators of quantum observables is then carried out and yields explicit expressions for quantum corrections.  
We also provide the detailed analysis of quantum integrability for a wide class 
of examples within the above list.

The concluding section 
includes a discussion and brings together several remarks about the status of 
the ``minimal'' quantization that has been abstracted from the various examples 
dealt with in this paper. It also opens some prospects for future investigations 
related to quantum integrability in the spirit of this work.

\bigskip
\textbf{Acknowledgements}:
We are indebted to Daniel Bennequin for several very interesting remarks, and to Brandon Carter for fruitful correspondence. Special thanks are due to Valentin Ovsienko for a careful reading of the manuscript together with a number of most enlightening suggestions.


\section{Classical integrable systems}\label{ClassIntSys}


\subsection{Killing tensors}\label{KT}

Let us start with the definition of the Schouten bracket of two polynomial functions
on the cotangent bundle $(T^*M,\omega=d\xi_i\wedge{}dx^i)$ of a smooth manifold~$M$. 
Consider two such homogeneous 
polynomials  $P=P^{i_1\ldots{}i_k}(x)\xi_{i_1}\ldots\xi_{i_k}$ and
$Q=Q^{i_1\ldots{}i_\ell}(x)\xi_{i_1}\ldots\xi_{i_\ell}$ of degree~$k$ and $\ell$ respectively; we will identify these 
polynomials with the corresponding smooth symmetric contravariant tensor fields $P^\sharp=P^{i_1\ldots{}i_k}(x)\partial_{i_1}\otimes\cdots\otimes\partial_{i_k}$ and
$Q^\sharp=Q^{i_1\ldots{}i_\ell}(x)\partial_{i_1}\otimes\cdots\otimes\partial_{i_\ell}$.

The Schouten bracket $[P^\sharp,Q^\sharp]_S$ of the two contravariant symmetric 
tensors $P^\sharp$ and $Q^\sharp$ (of degree $k$ and $\ell$ respectively) is the 
symmetric contravariant $(k+\ell-1)$-tensor 
corresponding to the Poisson bracket of $P$ and $Q$, namely 
\begin{equation}
[P^\sharp,Q^\sharp]_S=\{P,Q\}^\sharp.
\label{DefSchouten}
\end{equation}

Using the the Poisson bracket
$\{P,Q\}=\partial_{\xi_i}P\partial_iQ-\partial_{\xi_i}Q\partial_iP$, and 
(\ref{DefSchouten}), we readily get the local expression of the Schouten bracket 
of $P^\sharp$ and $Q^\sharp$. 
If the manifold $M$ is endowed with a symmetric connection $\nabla$, the latter 
can be written as\footnote{In this article
the round (resp. square) brackets will denote symmetrization (resp. skew-symmetrization) 
with the appropriate combinatorial factor.}
\begin{equation}
[P^\sharp,Q^\sharp]_S^{i_1\ldots{}i_{k+\ell-1}}
=
k\,P^{i(i_1\ldots{}i_{k-1}}\nabla_iQ^{i_k\ldots{}i_{k+\ell-1})}
-
\ell\,Q^{i(i_1\ldots{}i_{\ell-1}}\nabla_iP^{i_{\ell}\ldots{}i_{k+\ell-1})}.
\label{SchoutenNabla}
\end{equation}

If $M$ is, in addition, equipped with a (pseudo-)Riemannian metric,~$\rg$, we denote
by 
\begin{equation}
H=\half\,\rg^{ij}\xi_i\xi_j
\label{H}
\end{equation}
the Hamiltonian function associated with this structure. The Hamiltonian flow 
associated with $H$ is nothing but the geodesic flow on $T^*M$. 

A symmetric contravariant tensor field
$P^\sharp$ of degree $k$ satisfying
$
\{H,P\}=0
$
is called a Killing (or Killing-St\"ackel) tensor; using now the Levi-Civita connection $\nabla$ in  (\ref{SchoutenNabla}), this condition reads
\begin{equation}
\nabla^{(i}P^{i_1\ldots{}i_k)}=0.
\label{DefKilling}
\end{equation}


\goodbreak

\subsection{Killing-Maxwell tensors}\label{MaxwellKilling}

\subsubsection{Souriau's coupling}

In the presence of an electromagnetic field, $F$, Souriau \cite{Sou} has proposed to 
replace the canonical symplectic structure,~$\omega$, of
$T^*M$ by the twisted symplectic structure
$\omega_F=d\xi_i\wedge{}dx^i+\half{}F_{ij}dx^i\wedge{}dx^j$. The (gauge-invariant) 
Poisson bivector now reads
$$
\pi_F=\partial_{\xi_i}\wedge\partial_i-\half{}F_{ij}\,
\partial_{\xi_i}\wedge\partial_{\xi_j}.
$$
The Poisson bracket of two observables $P,Q$ of $T^*M$ is now 
\begin{equation}
\{P,Q\}_F=\pi_F(dP,dQ)
=
\partial_{\xi_i}P\partial_iQ-\partial_{\xi_i}Q\partial_iP-
F_{ij}\,\partial_{\xi_i}P\wedge\partial_{\xi_j}Q,
\label{PBF}
\end{equation}
and the Schouten-Maxwell bracket of two polynomials $P$ and $Q$ is then defined by
$$
[P^\sharp,Q^\sharp]_{S,F}=\{P,Q\}^\sharp_F.
$$

If the manifold $M$ is endowed  with a symmetric connection $\nabla$, the 
Schouten-Maxwell bracket takes on the following form
\begin{equation}
\begin{array}{rcl}
[P^\sharp,Q^\sharp]_{S,F}
=&&
[P^\sharp,Q^\sharp]_S^{i_1\ldots{}i_{k+\ell-1}}\partial_{i_1}
\otimes\cdots\otimes\partial_{i_{k+\ell-1}}\\[6pt]
&&
-k\ell\,F_{ij}P^{i(i_1\ldots{}i_{k-1}}\,Q^{i_k\ldots{}i_{k+\ell-2})j}
\partial_{i_1}\otimes\cdots\otimes\partial_{i_{k+\ell-2}}
\end{array}
\label{SchoutenMaxwell}
\end{equation}
with the expression (\ref{SchoutenNabla}) of the Schouten bracket $[\,\cdot\,,\,\cdot\,]_S$.

\goodbreak

Suppose now that the manifold $M$ is endowed with a metric $\rg$; the Hamiltonian vector 
field on $(T^*M,\omega_F)$ for the Hamiltonian $H$ given by 
(\ref{H}) yields the the Lorentz equations of motions for a charged test particle moving 
on $(M,\rg)$ under the influence of an external electromagnetic field $F$.

A symmetric contravariant tensor field
$P^\sharp$ of degree $k$ on $(M,\rg)$
is now called a Killing-Maxwell tensor if
$\{H,P\}_F=0$.
The Killing-Maxwell equations then read, using~(\ref{SchoutenMaxwell}), 
\begin{equation}
\nabla^{(i}P^{i_1\ldots{}i_k)}=0
\qquad
\&
\qquad
P^{i(i_1\ldots{}i_{k-1}}\,F^{i_k)}_i=0
\label{KillingMaxEqs}
\end{equation}
where $F^j_i=\rg^{jm}F_{mi}$, in accordance with previous results \cite{BCar3} 
obtained with a slightly different standpoint. 

The conditions (\ref{KillingMaxEqs}) are of special importance for proving the classical and quantum integrability of the equations of motion of a charged test particle in the generalized Kerr-Newman background.

\subsubsection{Standard electromagnetic coupling}

A more traditional, though equivalent, means to deal with the coupling to an electromagnetic field, 
$F=dA$ (locally), is to keep the canonical $1$-form, $\alpha=\xi_i dx^i$, on $T^*M$ 
unchanged, and hence to work with the original Poisson bracket $\{\cdot,\cdot\}$, but 
to replace the Hamiltonian (\ref{H}) by
\begin{equation}
\widetilde{H}=\half\rg^{ij}(\xi_i-A_i)(\xi_j-A_j)
\label{HF}
\end{equation}
where the tilde makes it clear that the expressions to consider are actually polynomials 
in the variables $\xi_i-A_i$, for $i=1,\ldots,n$; for example, if $P=P^{i_1\cdots{}i_k}\xi_{i_1}\ldots\xi_{i_k}$, then
\begin{equation}
\widetilde{P}=P^{i_1\ldots{}i_k}(\xi_{i_1}-A_{i_1})\ldots(\xi_{i_k}-A_{i_k}).
\label{tilde}
\end{equation}

The equations of motion given by the Hamiltonian vector field for the Hamiltonian 
(\ref{HF}) on $(T^*M,d\alpha)$ are, again, the Lorentz equations of motion.

The Schouten-Maxwell brackets and Schouten brackets for the electromagnetic coupling 
are related as follows via the corresponding Poisson brackets, viz
$$
\{P,Q\}_{F}=\{\widetilde{P},\widetilde{Q}\}.
$$

In this framework, a Killing-Maxwell tensor, $P^\sharp$, of degree $k$ on $(M,\rg)$ is 
defined by the equation $\{\widetilde{H},\widetilde{P}\}=0$. The resulting constraints 
are, again, given by (\ref{KillingMaxEqs}).

{}From now on, and in order to simplify the notation, we will omit the $\sharp$-superscript 
and use the same symbol for symmetric contravariant tensors and the corresponding 
polynomial functions on $T^*M$.


\subsection{General definition of classical integrability}

Let us recall that a dynamical system $(\cM,\omega,H)$ is (Liouville) integrable if there exist 
$n=\half\dim{\cM}$ independent Poisson-commuting functions 
$P_1,\ldots,P_n\in{}C^\infty(\cM)$ --- that is $dP_1\wedge\cdots\wedge{}dP_n\neq0$ 
and $\{P_k,P_\ell\}=0$ for all $k,\ell=1,\ldots,n$ --- such that
$P_1=H$.

We will, in the sequel, confine considerations to the case of cotangent bundles, 
$(\cM=T^*M,\omega=d\theta)$ where $\theta$ is the canonical $1$-form, and of 
polynomial functions, $P_1,\ldots,P_n$, on $T^*M$, that is to the case of $n$ 
Schouten-commuting Killing tensors. 
Moreover, all examples that we will consider will be given by 
polynomials of degree two or three.

\subsection{The St\"ackel systems}\label{StaeckelSection}

These systems on $(T^*M,\omega=d\xi_i\wedge{}dx^i)$ are governed by 
the Hamiltonians
\begin{equation}
H=\sum_{i=1}^n a^i(x) \left(\rule{0mm}{5mm}\half\xi_i^2+f_i(x^i)\right)
\label{HStaeckel}
\end{equation}
where the $i$-th function $f_i$ depends on the coordinate $x^i$ only, and  
the functions $a^i$ are defined as follows. Let $B$ denote a $\GL(n,\bR)$-valued 
function defined on $M$ and such that 
$$
B(x)=(B_1(x^1)B_2(x^2)\ldots{}B_n(x^n))
$$ 
where the $i$-th column $B_i(x^i)$ depends on $x^i$ only ($i=1,\ldots,n$); such 
a matrix will be called a St\"ackel matrix. Then take 
$$
a(x)=\pmatrix{a^1(x)\cr\vdots\cr{}a^n(x)}
$$
to be the first column $A_1(x)$ of the matrix $A(x)=B(x)^{-1}$. 

The integrability of such a system follows from the existence of 
$n$ quadratic polynomials
\begin{equation}\label{Staeckelconserved}
I_\ell=\sum_{i=1}^n A_\ell^i(x)\left(\rule{0mm}{5mm}\half\xi_i^2+f_i(x^i)\right),\qquad 
\ell=1,\ldots, n,\qquad H=I_1.
\end{equation}
We call St\"ackel potential every function of the form 
\begin{equation}\label{StaeckelPotential}
U_\ell(x)=\sum_{i=1}^n{A_\ell^i(x)f_i(x^i)},\qquad \ell=1,\ldots,n;
\end{equation}
the potential appearing in the Hamiltonian is just $U_1.$

One can check (see, e.g., \cite{Per}, p. 101) that the $n$ independent 
quantities $I_\ell$ are such that
\[
\{I_\ell,I_m\}=\sum_{s,t=1}^n\,(A_\ell^s\,\partial_s A_m^t-A_m^s\,\partial_s A_\ell^t)
\xi_s\left(\half\xi_t^2+f_t\right),\qquad \ell\neq{}m.
\]
The relation $A=B^{-1}$, gives the useful identity\footnote{The Einstein summation 
convention is not used.}
\begin{equation}\label{idStaeckel}
\partial _k A^i_j=-C^i_k\,A_j^k,\qquad\quad C^i_k=\sum_{s=1}^nA^i_s\frac{dB^s_k}{dx^k},
\end{equation}
which implies
\begin{equation}\label{idStaeckelbis}
A_\ell^s\,\partial_s A_m^t-A_m^s\,\partial_s A_\ell^t=0,\quad \ell\neq{}m,\quad s,t=1,\ldots,n
\end{equation}
and therefore the so defined St\"ackel systems are classically integrable.
\begin{rmk}\label{RmkPars}
{\rm
Let us mention an interesting result due to Pars (see \cite{Per}, p.~102): for a system whose Hamiltonian is of the form (\ref{HStaeckel}), the Hamilton-Jacobi equation is separable if and only if this system is St\"ackel.
}
\end{rmk}
Although these systems constitute quite a large class of integrable 
systems, they do not exhaust the full class. A simple example 
of a non-St\"ackel integrable system was produced by Di Pirro 
(see Section \ref{ClassicalDiPirro}).

\subsection{The Jacobi integrable system on the ellipsoid}
\label{subellipsoid}

Let $\cE\subset\bR^{n+1}$ be the $n$-dimensional ellipsoid defined by the equation $Q_0(y,y)=1$ where 
we define, for $y,z\in\bR^{n+1}$,
\begin{equation}
Q_\lambda(y,z)
=
\sum_{\alpha=0}^n{\frac{y_\alpha{}z_\alpha}{a_\alpha-\lambda}},
\label{Qlambda}
\end{equation}
with $0<a_0<a_1<\ldots<a_n$; the equations $Q_\lambda(y,y)=1$ define a family of 
confocal quadrics. 

\goodbreak

It has been proved by Jacobi (in the case $n=2$) that the differential equations governing the geodetic motions on the ellipsoid, $\cE$, form an integrable system. The same remains true if a quadratic potential is admitted (see \cite{Mos2}). The Hamiltonian of the system, prior to reduction, reads
\begin{equation}
H(p,y)=\frac{1}{2}\sum_{\a=0}^{n}{p_\a^2}+\frac{a}{2}\sum_{\a=0}^{n}{y_\a^2}
\label{HamiltonianJacobi}
\end{equation}
where $p,y\in\bR^{n+1}$ and $a$ is some real parameter.

Moser has shown \cite{Mos1} that the following polynomial functions
\begin{equation}
F_\alpha(p,y)=p_\alpha^2+ay_\a^2+\sum_{\beta\neq\alpha}{
\frac{\left(p_\alpha{}y_\beta-p_\beta{}y_\alpha\right)^2}{a_\alpha-a_\beta}
}
\qquad
\mathrm{with}
\qquad
\alpha=0,1,\ldots,n,
\label{IntegralsIninvolutionEllipsoid}
\end{equation}
are in involution on $(T^*\bR^{n+1},\sum_{\a=0}^n{dp_\alpha\wedge{}dy_\alpha})$. Those will 
generate the commuting first integrals of the Jacobi dynamical system on the cotangent 
bundle $T^*\cE$ of the ellipsoid.

Our goal is to deduce from the knowledge of (\ref{IntegralsIninvolutionEllipsoid}) 
the independent quantities in involution $I_1,\ldots,I_n$ on 
$(T^*\cE,d\xi_i\wedge{}dx^i)$ from the symplectic embedding 
$$
\iota:T^*\cE\hookrightarrow{}T^*\bR^{n+1}
$$ 
given by $Z_1(p,y)=Q_0(y,y)-1=0$ and $Z_2(p,y)=Q_0(p,y)=0$. 

\begin{pro}
The restrictions $F_\alpha\big{\vert}_{T^*\cE}=F_\alpha\circ\iota$ 
of the functions (\ref{IntegralsIninvolutionEllipsoid}) Poisson-commute on $T^*\cE$. 
\end{pro}
\begin{proof}
We get, using Dirac brackets,
\begin{equation}\label{Diracbracket}\begin{array}{l}
\{F_\a\big{\vert}_{T^*\cE},F_\b\big{\vert}_{T^*\cE}\}=
\{F_\a,F_\b\}\big{\vert}_{T^*\cE}\\[6pt]
\displaystyle\hspace{3cm} -
\frac{1}{\{Z_1,Z_2\}}
\left[
\{Z_1,F_\a\}\{Z_2,F_\b\}-\{Z_1,F_\b\}\{Z_2,F_\a\}
\right]\big{\vert}_{T^*\cE}
\end{array}\end{equation}
for second-class constraints. Now, the denominator
$\{Z_1,Z_2\}=-2\sum_{\a=0}^n{(y_\a/a_\a)^2}$ doesn't vanish while
$\{Z_1,F_\alpha\}=4(p_\alpha y_\alpha/a_\alpha) Z_1-4 (y_\alpha^2/a_\alpha)Z_2$ 
is zero on~$T^*\cE$, for all $\alpha=0,...,n$. The fact that $\{F_\a,F_\b\}=0$ 
completes the proof.
\end{proof}

The reduced Hamiltonian for the Jacobi system on the ellipsoid $\cE$ is 
plainly 
\begin{equation}
H=\frac{1}{2}\,\sum_{\alpha=0}^n{\left(p_\alpha^2+a{}y_\a^2\right)}\Big{\vert}_{T^*\cE}
=
\frac{1}{2}\,\sum_{\alpha=0}^n{F_\alpha\Big{\vert}_{T^*\cE}}.
\label{Hconstrained}
\end{equation}

In order to provide explicit expressions for the function in involution 
$I_1,\ldots,I_n$, we resort to Jacobi 
ellipsoidal coordinates $x^1,\ldots,x^n$ on $\cE$. Those are defined by 
\begin{equation}
Q_\lambda(y,y)=1-\frac{\lambda{}U_x(\lambda)}{V(\lambda)}
\label{EllipsoidalCoordinates}
\end{equation}
where
\begin{equation}
U_x(\lambda)=\prod_{i=1}^n(\lambda-x^i)
\qquad
\mathrm{and}
\qquad
V(\lambda)=\prod_{\alpha=0}^n(\lambda-a_\alpha)
\label{UV}
\end{equation}
and are such that $a_0<x^1<a_1<x^2<\ldots<x^n<a_n$.
The induced metric, $\rg=\sum_{i,j=1}^n{\rg_{ij}(x)dx^idx^j}$, of the ellipsoid $\cE$ 
is given by
$$
\rg_{ij}(x)=\frac{1}{4}\sum_{\alpha=0}^n{
\frac{y_\alpha^2}{(a_\alpha-x^i)(a_\alpha-x^j)}
}
$$
and retains the form \cite{Mos1}
\begin{equation}
\rg=\sum_{i=1}^n\rg_{i}(x)(dx^i)^2
\qquad
\mathrm{where}
\qquad
\rg_i(x)=-\frac{x^i}{4}\frac{U'_x(x^i)}{V(x^i)}
\label{MetricEllipsoid}
\end{equation}
which is actually Riemannian because of the previous inequalities. We put for 
convenience $\rg^i(x)=1/\rg_i(x)$.

Using (\ref{EllipsoidalCoordinates}) and (\ref{UV}), we find the local expressions $y_\alpha(x)$ via the formula
\begin{equation}
y_\alpha^2
=
a_\alpha
\frac{\displaystyle
\prod_{i=1}^n{(a_\alpha-x^i)}
}{\displaystyle
\prod_{\beta\neq\alpha}{(a_\alpha-a_\beta)}
}
\label{y2}
\end{equation}
and then obtain the constrained coordinate functions
\begin{equation}
p_\alpha(\xi,x)=-\half{}y_\alpha\sum_{i=1}^n{
\frac{\rg^i(x)\xi_i}{(a_\alpha-x^i)}
}
\label{palpha}
\end{equation}
given by the induced canonical $1$-form 
$\sum_{i=1}^n{\xi_i\,dx^i}=\iota^*\left(\sum_{\alpha=0}^n{p_\alpha{}dy_\alpha}\right)$.

The Hamiltonian (\ref{Hconstrained}) on $(T^*\cE,d\xi_i\wedge{}dx^i)$ is then found to be
\begin{equation}
H=
\frac{1}{2}\sum_{i=1}^n{\rg^i(x)\xi_i^2}
+
\frac{a}{2}\left[\sum_{\a=0}^n{a_\alpha}-\sum_{i=1}^n{x^i}\right].
\label{Hellipsoid}
\end{equation}
Note that the potential term is obtained from the large $\lambda$ behaviour 
\[Q_{\lambda}(y,y)\sim\frac 1{\lambda} \sum_{\alpha=0}^ny_{\alpha}^2+
\frac 1{\lambda^2}\sum_{\alpha=0}^n a_{\alpha}y_{\alpha}^2+\cdots\]
which can be computed using relation (\ref{EllipsoidalCoordinates}). One gets
\[Q_{\lambda}(y,y)\sim\frac 1{\lambda}\left[\rule{0mm}{4mm}
\sum_{\a=0}^n{a_\alpha}-\sum_{i=1}^n{x^i}\right]+\cdots\]

One relates the conserved quantities (\ref{IntegralsIninvolutionEllipsoid}) to 
their reduced expressions on~$T^*\cE$ by computing, using (\ref{palpha}) 
and (\ref{y2}), the expression of $F_\alpha\big{\vert}_{T^*\cE}$. One gets the

\goodbreak

\begin{pro} 
The Moser conserved quantities $\left(F_\alpha\big{\vert}_{T^*\cE}\right)_{\alpha=0,\ldots,n}$ 
retain the form
$$
F_\alpha\big{\vert}_{T^*\cE}
=
\frac{a_\alpha\,G_{a_\alpha}(\xi,x)}{\displaystyle
\prod_{\beta\neq\alpha}{(a_\alpha-a_\beta)}
}
$$
where
\begin{equation}\label{GenFun}
G_\lambda(\xi,x)
=
\sum_{i=1}^n{
g^i(x)\prod_{j\neq{}i}{(\lambda-x^j)\xi_i^2}
}
+
a\prod_{i=1}^n{(\lambda-x^i)}.
\end{equation}
\end{pro}

It is useful to introduce the notation $\sigma_k^i(x)$ for the symmetric 
functions of order $k=0,1,\ldots,n-1$ of the variables $(x^1,\ldots,x^n)$, 
with the exclusion of index $i$, namely
\begin{equation}\label{fctgene}
\prod_{j\neq{}i}(\lambda-x^j)
=
\sum_{k=1}^n{(-1)^{k-1}
\lambda^{n-k}\sigma_{k-1}^i(x)
}.
\end{equation}
We note that, from the above definition, $\sigma_0^i(x)=1$.

It is also worthwhile to introduce other symmetric functions, $\s_k(x)$, via
\begin{equation}\label{fctgene2}
\prod_{j=1}^n(\lambda-x^j)
=
\sum_{k=0}^n{(-1)^{k}
\lambda^{n-k}\sigma_{k}(x)
}.
\end{equation}

We thus have
\begin{equation}
G_\lambda(\xi,x)
=
\sum_{i=1}^n{(-1)^{i-1}
\lambda^{n-i}I_i(\xi,x)
}
+a(-\lambda)^n
\label{GJacobi}
\end{equation}
where the independent functions $I_i$ ($i=1,\ldots,n$) are in involution 
and can be written as
\begin{equation}
I_i(\xi,x)=\sum_{j=1}^n{
A_i^j(x)\xi_j^2-a\sigma_i(x)
}
\qquad
\mathrm{with}
\qquad
A_i^j(x)=g^j(x)\sigma_{i-1}^j(x).
\label{PiEllipsoid}
\end{equation}
In the case $i=1$, we recover the Hamiltonian (\ref{Hellipsoid}), i.e., 
$$
H=\frac{1}{2}\,I_1+\frac{a}{2}\sum_{\a=0}^n{a_\a}.
$$ 

\goodbreak

\begin{pro}\label{ProJacobi}
The Jacobi system on $T^{\star}{\cal E}$ defines a St\" ackel system, with 
St\" ackel matrix
\begin{equation}
B^i_k(x^k)=(-1)^i\,\frac{(x^k)^{n+1-i}}{4 V(x^k)}
\label{BN}
\end{equation}
and potential functions
\begin{equation}
f_k(x^k)=a\frac{(x^k)^{n+1}}{4 V(x^k)}
\label{fk}
\end{equation}
for $i,k=1,\ldots,n$.
\end{pro}
\begin{proof}
It is obvious from its definition that $B$ is a St\" ackel matrix. We just need 
to prove that $A=B^{-1}.$ To this aim we first prove a useful identity. 
Let us consider the integral in the complex plane
\[
\frac 1{2i\pi}\int_{\vert{z}\vert=R}\frac{z^{n-i}}{(z-\lambda)}
\frac{U_x(\lambda)}{U_x(z)}\,dz.
\]
When $R\to\infty$ 
the previous integral vanishes because the integrand vanishes as $1/R^2$ for large 
$R.$ We then compute this integral using the theorem of residues and we get 
the identity
\begin{equation}\label{usefulidentity}
\sum_{k=1}^n\frac{(x^k)^{n-i}}{U'_x(x^k)}\ \prod_{j\neq k}(\lambda-x^j)=\lambda^{n-i}.
\end{equation}
Equipped with this identity let us now prove that
\[\sum_{k=1}^n B^i_k\,A^k_j=\delta^i_j.\]
Multiplying this relation by $(-1)^{j-1}\lambda^{n-j}$ and summing over $j$ from $1$ 
to $n$, we get the equivalent relation
\[\sum_{k=1}^n B^i_k\,\sum_{j=1}^n (-1)^{j-1}\lambda^{n-j}A^k_j= 
(-1)^{i-1}\lambda^{n-i},\]
which becomes, using (\ref{PiEllipsoid}) and (\ref{fctgene}):
\[\sum_{k=1}^n B^i_k\,g^k(x)\prod_{j\neq k}(\lambda-x^j)=(-1)^{i-1}\lambda^{n-i}.\]
Using the explicit form of $g^k(x)$ given in (\ref{MetricEllipsoid}) and of the matrix 
$B$, this relation reduces to the identity (\ref{usefulidentity}) and this completes 
the derivation of (\ref{BN}).

In order to get the functions $f_i(x^i)$ as in (\ref{HStaeckel}), let us resort to (\ref{PiEllipsoid}) and solve, for the unknown $f_i$, the following equation
$$
-a\sigma_i(x)=\sum_{j=1}^n{A_i^j(x)f_j}.
$$
Multiplying both sides by $B_k^i$, summing over $i$ from $1$ to $n$, and using (\ref{BN}) we get
\begin{eqnarray*}
f_k
&=&
-a\sum_{i=1}^n{B^i_k\sigma_i(x)}=
-\frac{a}{4 V(x^k)}\sum_{i=1}^n{(-1)^i(x^k)^{n+1-i}\sigma_i(x)}\\
&=&
-\frac{a}{4 V(x^k)}\left[\sum_{i=0}^n{(-1)^i(x^k)^{n+1-i}\sigma_i(x)}-(x^k)^{n+1}\right].
\end{eqnarray*}
In view of (\ref{fctgene2}), we have $\sum_{i=0}^n{(-1)^i(x^k)^{n-i}\sigma_i(x)}=\prod_{j=1}^n(x^k-x^j)=0$, which completes the proof.
\end{proof}

\begin{rmk}\label{potJacobi}
{\rm
\begin{enumerate}
\item 
The fact that the geodesic flow on $T^{\star}{\cal E}$ is a St\" ackel 
system was first proved by Benenti in \cite{Ben}. 
We have given here a new derivation, which makes the link between Moser's conserved quantities on $T^*\bR^{n+1}$ and the St\"ackel conserved quantities on $T^*\cE$. We have extended this link to the case where Jacobi's potential is admitted.

\item 
Checking that 
the unconstrained observables $I_i$ are in involution is most 
conveniently done using their generating function (\ref{GenFun}). Indeed it is 
easy to verify the relation
\[ \{G_{\lambda}(\xi,x),G_{\mu}(\xi,x)\}=0,\qquad \lambda,\ \mu\in\bR,\]
which implies, via (\ref{GJacobi}), and upon expansion in powers of $\lambda$ and $\mu,$ the  relations $\,\{I_i,I_j\}=0$ for any $i,j=1,\ldots,n$.

\item 
Some authors \cite{BT,HW} have quantized the full set of commuting observables for the geodesic flow of the ellipsoid $\cE\subset\bR^{n+1}$ in its unconstrained form, namely on $T^*\bR^{n+1}$. Notice though that in the reduction process from $T^*\bR^{n+1}$ to $T^*\cE$ quantum corrections may prove necessary in order to insure self-adjointness of the quantized observables. Our point of view will be to perform the classical reduction in the first place and then to quantize the observables directly on $T^*\cE$ via a specific procedure that will be described in Section \ref{QuantIntSys}.
\end{enumerate}
}
\end{rmk}

\subsection{The Neumann system}\label{subNeumann}

The Neumann Hamiltonian on $(T^*\bR^{n+1},\sum_{\a=0}^n{dp_{\alpha}\wedge dy_{\alpha}})$ is
\begin{equation}
H=\half\sum_{\a=0}^n\left(p_\a^2+a_\a y_\a^2\right)
\label{HNeumann}
\end{equation}
with the real parameters $0<a_0<a_1<\ldots<a_n.$ Under the symplectic reduction, 
with the second class constraints
\begin{equation}
Z_1(p,y)=\sum_{\a=0}^n y_\a^2-1=0,\qquad\qquad Z_2(p,y)=\sum_{\a=0}^n{p_\a y^\a}=0,
\label{NeumannConstraints}
\end{equation}
it becomes a dynamical system on $(T^*S^n,d\xi_i\wedge{}dx^i)$.

This system is classically integrable, with the following commuting first integrals 
of the Hamiltonian flow in $T^*\bR^{n+1}$:
\begin{equation}
F_\alpha(p,y)=y_\alpha^2+\sum_{\beta\neq\alpha}{
\frac{\left(p_\alpha{}y_\beta-p_\beta{}y_\alpha\right)^2}{a_\alpha-a_\beta}
}
\qquad
\mathrm{with}
\qquad
\alpha=0,1,\ldots,n.
\label{IntegralsIninvolutionNeumann}
\end{equation}

The symplectic embedding 
$$
\iota:T^* S^n\hookrightarrow{}T^*\bR^{n+1}
$$ 
given by $Z_1(p,y)=0$ and $Z_2(p,y)=0$ preserves the previous conservation laws. 
Indeed  the Poisson brackets of the restrictions 
$F_\alpha\big{\vert}_{T^*\cE}=F_\alpha\circ\iota$ of the functions $F_{\alpha}$ 
are still given by the Dirac brackets 
(\ref{Diracbracket}) of the second class constraints (\ref{NeumannConstraints}). This time we have
\[\{Z_1,Z_2\}=-2\sum_{\alpha=0}^n y_{\alpha}^2\neq 0,\qquad \{Z_1,F_{\alpha}\}=0,\]
which gives again
\[\{F_\a\big{\vert}_{T^*\cE},F_\b\big{\vert}_{T^*\cE}\}=0.\]

Let us introduce an adapted coordinate system on $(T^*S^n,d\xi_i\wedge{}dx^i)$ much 
in the same manner as for the ellipsoid.

We start with the following definition \cite{Mos1} of a coordinate system 
$(x^1,\ldots,x^n)$ on $S^n$:
$$
Q_\l(y,y)
=
\sum_{\a=0}^n{\frac{y_\a^2}{a_\a-\l}}
=
-\frac{\prod_{i=1}^n(\l-x^i)}{\prod_{\a=0}^n(\l-a_\a)}.
$$
The following inequalities hold: $0<a_0<x^1<a_1<\ldots<x^n<a_n$.
We get, in the same way as before,
\begin{equation}
y_\a^2=\frac{\prod_{i=1}^n(a_\a-x^i)}{\prod_{\b\neq\a}(a_\a-a_\b)}
\label{y2Neumann}
\end{equation}
together with the following expression of the round metric 
$\rg=\sum_{\a=0}^n dy_\a^2\big\vert_{S^n}$ in terms of the newly introduced coordinates, namely
\begin{equation}\label{metricNeumann}
\rg=\sum_{i=1}^n\rg_i(x)(dx^i)^2
\qquad
\mathrm{with}
\qquad
\rg_i(x)=-\frac{U'_x(x^i)}{4 V(x^i)}
\end{equation}
with the notation (\ref{UV}). Again, we put for convenience $\rg^i(x)=1/\rg_i(x)$.

Our goal is to deduce from the knowledge of (\ref{IntegralsIninvolutionNeumann}) 
the independent quantities in involution $I_1,\ldots,I_n$ on 
$(T^*S^n,d\xi_i\wedge{}dx^i)$. The formula (\ref{palpha}) relating unconstrained 
and constrained momenta still holds and yields the
\begin{pro} 
The Neumann system $\left(F_\alpha\big{\vert}_{T^*S^n}\right)_{\alpha=0,\ldots,n}$ 
retains the following form
$$
F_\alpha\big{\vert}_{T^*S^n}
=
-\frac{\,G_{a_\alpha}(\xi,x)}{\displaystyle
\prod_{\beta\neq\alpha}{(a_\alpha-a_\beta)}
}
$$
where
$$
G_\lambda(\xi,x)
=
\sum_{i=1}^n{
g^i(x)\prod_{j\neq{}i}{(\lambda-x^j)\xi_i^2}
}
+
\prod_{j=1}^n(\l-x^j).
$$
\end{pro}

Let us, again, posit
$$
G_\lambda(\xi,x)
=
\sum_{i=1}^n{(-1)^{i-1}
\lambda^{n-i}I_i(\xi,x)
}
+\l^n
$$
where the independent functions $I_i$ ($i=1,\ldots,n$) are in involution 
and can be written as
\begin{equation}
I_i(\xi,x)=\sum_{j=1}^n{
A_i^j(x)\xi_j^2
}
-\s_i(x)
\qquad
\mathrm{with}
\qquad
A_i^j(x)=g^j(x)\sigma_{i-1}^j(x),
\label{PiNeumann}
\end{equation}
where the symmetric functions $\s_i(x)$ are as in (\ref{fctgene2}).

Using the relations 
$$
\s_1(x)=\sum_{i=1}^n x^i,
\qquad
\mathrm{and}
\qquad
\sum_{\a=0}^n{a_\a y_\a^2}=\sum_{\a=0}^n a_\a-\sum_{i=1}^n x^i,
$$
one can check that the Hamiltonian (\ref{HNeumann}) is $H=\half{}I_1$.

\begin{pro} The Neumann flow on $(T^*S^n,H)$ defines a St\" ackel system, with 
St\" ackel matrix
$$
B^i_k(x^k)=(-1)^i\,\frac{(x^k)^{n-i}}{4 V(x^k)}
$$
and potential functions
\begin{equation}
f_k(x^k)=\frac{(x^k)^{n}}{4 V(x^k)}
\label{fkbis}
\end{equation}
for $i,k=1,\ldots,n$.
\end{pro}
\begin{proof}
To check that $A=B^{-1}$, it is enough to use the identity (\ref{usefulidentity}). The computation of the potential functions $f_k$ proceeds along the same lines as in the proof of Proposition \ref{ProJacobi}.
\end{proof}

\begin{rmk}
{\rm
The involution property $\{I_i,I_j\}=0$ for $i,j=1,\ldots,n$, similarly to the case of the 
ellipsoid, is seen to follow from the relation 
$\{G_{\lambda}(\xi,x),G_{\mu}(\xi,x)\}=0.$
}
\end{rmk}

\subsection{Test particles in generalized Kerr-Newman background}\label{KNS}

Plebanski and Demianski have constructed in \cite{Ple,PD} a class of metrics generalizing the Kerr-Newman solution in $4$-dimensional spacetime. The former are also known as the Kerr-Newman-Taub-NUT-de Sitter solutions of the Einstein-Maxwell equations. The metric, in the coordinate system 
$(x^1,x^2,x^3,x^4)=(p,q,\sigma,\tau)$, retains the form
\begin{equation}
\rg=\frac{X}{p^2+q^2}(d\tau+q^2 d\sigma)^2
-\frac{Y}{p^2+q^2}(d\tau-p^2 d\sigma)^2
+\frac{p^2+q^2}{X}\,dp^2
+\frac{p^2+q^2}{Y}\,dq^2
\label{g}
\label{KNSmetric}
\end{equation}
with
\begin{equation}
X=\gamma-g^2+2np-\epsilon p^2-\frac{\Lambda}{3}\,p^4,
\qquad
\&
\qquad
Y=\gamma+e^2-2mq+\epsilon q^2-\frac{\Lambda}{3}\,q^4,
\label{XY}
\end{equation}
where $(m,\gamma)$ are related to the mass and angular momentum of the Kerr black hole, $(e,g)$ to the electric and magnetic charge; $n$ is the NUT charge, and $\Lambda$ the cosmological constant. The remaining parameter $\epsilon$ can be scaled out to $\pm1$ or $0$.

This metric, $\rg$, together with the electromagnetic field, locally given by 
$F=dA$ where
\begin{equation}
A=\frac{1}{p^2+q^2}\Big[
(e q + g p) d\tau + p q(g q - e p) d\sigma
\Big],
\label{A}
\end{equation}
provide an exact solution of the Einstein-Maxwell equations with cosmological 
constant $\Lambda$. Let us notice for further use that
\begin{equation}
\nabla_iA^i=0.
\label{DivA}
\end{equation}

Upon defining the $1$-forms
\begin{eqnarray*}
K&=&\sqrt{\frac{Y}{2(p^2+q^2)}}\,(d\tau-p^2\,d\sigma)+\sqrt{\frac{p^2+q^2}{2Y}}\,dq,\\
L&=&\sqrt{\frac{Y}{2(p^2+q^2)}}\,(d\tau-p^2\,d\sigma)-\sqrt{\frac{p^2+q^2}{2Y}}\,dq,\\
M_1&=&\sqrt{\frac{p^2+q^2}{X}}\,dp,\\
M_2&=&\sqrt{\frac{X}{p^2+q^2}}\,(d\tau+q^2 d\sigma),\\
\end{eqnarray*}
one constructs the $2$-form
\begin{equation}
\cY=
p K\wedge{}L-q M_1\wedge{}M_2.
\label{YanoKNS}
\end{equation}
One can check that the twice-symmetric tensor 
$P=-\cY^2$, namely $P_{ij}=-\cY_{ik}\cY_{\ell{}j}\rg^{k\ell}$, is a Killing-Maxwell 
tensor (see (\ref{KillingMaxEqs})), given by
\begin{equation}
P=
p^2(K\otimes{}L+L\otimes{}K)+q^2(M_1\otimes{}M_1+M_2\otimes{}M_2).
\label{KillingMaxwellKNS}
\end{equation}
We thus recover Carter's result \cite{BCar3} about the integrability of the Hamiltonian flow for a charged test particle in the generalized Kerr-Newman background in a different manner. 

\begin{rmk}
{\rm
The $2$-form $\cY$ in (\ref{YanoKNS}) defines what is usually called a Killing-Yano 
tensor \cite{GRvH,MCar}.
}
\end{rmk}

The four conserved quantities in involution for the generalized Kerr-Newman system are, respectively, 
\begin{equation}
\widetilde{H}=\half\,\rg^{ij}(\xi_i-A_i)(\xi_j-A_j), 
\qquad
\widetilde{P}=P^{ij}(\xi_i-A_i)(\xi_j-A_j)
\label{HP-KNS}
\end{equation}
 where $P$ is as in (\ref{KillingMaxwellKNS}), and
\begin{equation}
\widetilde{S}=\xi_3-A_3,
\qquad
\widetilde{T}=\xi_4-A_4.
\label{ST-KNS}
\end{equation}

\goodbreak

\subsection{The Multi-Centre geodesic flow}

The class of Multi-Centre Euclidean metrics in $4$ dimensions retain, in a local coordinate 
system $(x^i)=(t,(y^a))\in\bR\times\bR^3$, the form
\begin{equation}
\rg=\frac{1}{V(y)}(dt+A_a(y)dy^a)^2+V(y)\gamma
\label{mcm}
\end{equation}
with $\gamma=\d_{ab}\,dy^a dy^b$ the flat Euclidean metric in $3$-space, and 
$dV=\pm\star(dA)$ where~$\star$ is the Hodge star for $\gamma$. These conditions 
insure that the metric (\ref{mcm}) is Ricci-flat.

For some special potentials $V(y)$, the geodesic flow is integrable as shown in 
\cite{GR,CFH,Val}. The four conserved quantities in involution are given by 
\begin{equation}\label{Multicentre}
H=\half\rg^{ij}\xi_i\xi_j,\qquad K=K^i\xi_i,\qquad L=L^i\xi_i,\qquad P=P^{ij}\xi_i\xi_j,
\end{equation}
where $K$ and $L$ are two commuting Killing vectors and $P$ a Killing $2$-tensor whose expressions can be found in the previous References.
 
\subsection{The Di Pirro system}\label{ClassicalDiPirro}

Di Pirro has proved (see, e.g., \cite{Per}, p. 113)
that the Hamiltonian on $T^*\bR^3$
\begin{equation}
H=\frac{1}{2(\gamma(x^1,x^2)+c(x^3))}\left[a(x^1,x^2)\xi_1^2+b(x^1,x^2)\xi_2^2+
\xi_3^2\right]
\label{HDiPirro}
\end{equation}
admits one and only one additional first integral given by
\begin{equation}
P=\frac{1}{(\gamma(x^1,x^2)+c(x^3))}\left[c(x^3)\left(a(x^1,x^2)\xi_1^2+
b(x^1,x^2)\xi_2^2\right)-\gamma(x^1,x^2)\xi_3^2\right]. 
\label{PDiPirro}
\end{equation}

In the case where the metric defined by $H$ in (\ref{HDiPirro}) possesses a 
Killing vector, the system becomes integrable though not of St\"ackel type. 
This happens, e.g., if (i) $c(x^3)=\const$., or (ii) $a=b$ and $\gamma$ depend 
on $r=\sqrt{(x^1)^2+(x^2)^2}$ only.

\section{A quantization scheme for integrable systems}\label{QuantIntSys}

We wish to deal now with the quantum version of the preceding examples.
Let us start with some preliminary considerations:

\goodbreak

\begin{enumerate}
\item
There is no universally accepted procedure of quantization, i.e., of a linear 
identification, $\cQ$, of a space of classical observables with some space of 
linear symmetric operators on a Hilbert space. One --- among many --- of the pathways to construct such a 
quantization mapping has been to demand that the mapping $\cQ$ be equivariant 
with respect to some Lie group of symplecto\-morphisms of classical phase space. 
\item
Similarly, there is no universally accepted notion of quantum integrability. However, 
given a classical integrable system $P_1,\ldots,P_n$ on a symplectic manifold 
$(\cM,\omega)$, and a quantization mapping $\cQ:P_i\mapsto{}\widehat{P}_i$, we 
will say that such a system is integrable in the quantum sense if 
$[\widehat{P}_i,\widehat{P}_j]=0$ for all $i,j=1,\ldots,n$.
\item
A large number of integrable systems involve \textit{quadratic} observables. We 
will thus choose to concentrate on this important --- yet very special --- case, 
both from the classical and quantum viewpoint.
\item
Among all possible quantization procedures, the search for integrability-pre\-serving ones (if any) should be of fundamental importance. The quantization 
of quadratic observables we will present below might serve as a starting point for 
such a programme.
\end{enumerate}

\subsection{Quantizing quadratic and cubic observables}\label{SectionQuant}

Let us recall that the space $\cF_\l(M)$ of $\l$-densities on $M$ is defined as the 
space of sections of the complex line bundle 
$\left\vert\Lambda^n{}T^*M\right\vert^\l\otimes\bbC$. 
In the case where the configuration manifold is orientable, $(M,\vol)$, such a 
$\l$-density can be, locally, cast into the form $\phi=f\vert\vol\vert^\l$ 
with $f\in{}C^\infty(M)$ which means that $\phi$ transforms under the action 
of $a\in\Diff(M)$ according to $f\mapsto{}a_*f\vert(a_*\vol)/\vol\vert^\l$.

The completion $\cH(M)$ of the space of compactly supported half-densities, $\l=\half$, 
is a Hilbert space canonically attached to $M$ that will be used throughout this 
article. The scalar product of two half-densities reads
$$
\la\phi,\psi\ra=\int_M{\!\overline{\phi}\,\psi}
$$
where the bar stands for complex conjugation.

We will assume that configuration space is endowed with a (pseudo-)Riemannian structure, 
$(M,\rg)$; and denote by $\vert\vol_\rg\vert$ the corresponding density and by $\Gamma_{ij}^k$ the associated Christoffel symbols.

The quantization now introduced is a linear invertible mapping from the space of 
quadratic observables $P=P_2^{jk}(x)\xi_j\xi_k+P_1^j(x)\xi_j+P_0(x)$ to the space 
of second-order differential operators on $\cH(M)$, viz 
$A=\widehat{P}=A_2^{jk}(x)\nabla_j\nabla_k+A_1^j(x)\nabla_j+A_0(x)\bone$ where the 
covariant derivative of half-densities 
$\nabla_j\phi=\partial_j\phi-\half\Gamma_{jk}^k\phi$ (or, locally, 
$\nabla_j\phi=(\partial_jf)\vert\vol_\rg\vert^\half$) has been used. We furthermore 
require that the principal symbol be preserved (see below (\ref{PrincipalSymbol2}), (\ref{PrincipalSymbol1}) and (\ref{PrincipalSymbol0})), and that $\widehat{P}$ be 
formally self-adjoint, i.e., $\la\phi,\widehat{P}\psi\ra=\la\widehat{P}\phi,\psi\ra$ for all compactly supported $\phi,\psi\in\cF_\half(M)$.

The quantization reads
\begin{eqnarray}
\label{PrincipalSymbol2}
A_2^{jk}&=&-P_2^{jk}\\[6pt]
\label{PrincipalSymbol1}
A_1^{j}&=&iP_1^{j}-\nabla_kP_2^{jk}\\[6pt]
\label{PrincipalSymbol0}
A_0&=&P_0+\frac{i}{2}\nabla_jP_1^j
\end{eqnarray}
and admits the alternative form
\begin{equation}
\widehat{P}
=
-\nabla_j\scirc{}P_2^{jk}\scirc\nabla_k+\frac{i}{2}\left(P_1^j\scirc\nabla_j+
\nabla_j\scirc{}P_1^j\right)+P_0\bone
\label{Quantization012}
\end{equation}
which makes clear the symmetry of the quantum operators.

\begin{rmk}
{\rm
The formula (\ref{Quantization012}) was originally used by Carter \cite{BCar3} for 
proving the quantum integrability of the equations of motion of charged test particles in the Kerr-Newman solution.
}
\end{rmk}

\begin{rmk}
{\rm
It is worth mentioning that formula (\ref{Quantization012}) actually corresponds at the same time to the projectively equivariant quantization \cite{LO,DO2} and to the conformally 
equivariant quantization \cite{DLO,DO1} $\cQ_{0,1}(P):\cF_0(M)\to\cF_1(M)$ 
restricted to quadratic polynomials.
}
\end{rmk}

\goodbreak

One can check the relations:
\begin{equation}
[\widehat{P}_0,\widehat{Q}_1]
=
i[P_0,Q_1]_S
=
i\widehat{\{P_0,Q_1\}},
\label{CommP0Q1}
\end{equation}

\begin{equation}
[\widehat{P}_0,\widehat{Q}_2]
=
-\half\left(\nabla_j\scirc[P_0,Q_2]_S^j+[P_0,Q_2]_S^j\scirc\nabla_j
\right)
=
i\widehat{\{P_0,Q_2\}},
\label{CommP0Q2}
\end{equation}

\begin{equation}
[\widehat{P}_1,\widehat{Q}_1]
=
-\half\left(\nabla_j\scirc[P_1,Q_1]_S^j+[P_1,Q_1]_S^j\scirc\nabla_j
\right)
=
i\widehat{\{P_1,Q_1\}}.
\label{CommP1Q1}
\end{equation}

Quantum corrections appear explicitly whenever $k+\ell>2$, as can be seen from the 
next commutators:
\begin{equation}
[\widehat{P}_1,\widehat{Q}_2]
=
i\widehat{\{P_1,Q_2\}}
+
i\widehat{A}_{P_1,Q_2}
\label{CommP1Q2}
\end{equation}
where
\begin{equation}
A_{P_1,Q_2}=\frac{1}{2}\nabla_j \scirc Q_2^{jk}\scirc\nabla_k(\nabla_\ell P_1^\ell)
\label{A0}
\end{equation}
is a scalar quantum correction that may vanish in some special instances, e.g., if the 
vector-field~$P_1$ is divergence-free (in particular if it is a Killing vector-field).

The previous formul\ae\ can be found, in a different guise, in\cite{BCar3}. Here, we will go one step 
further and compute the commutators $[\widehat{P}_2,\widehat{Q}_2]$ which involve 
third-order differential operators. To that end, we propose to quantize homogeneous 
cubic polynomials according to
\begin{equation}
\widehat{P}_3=-\frac{i}{2}\left(
\nabla_j\scirc{}P_3^{jk\ell}\scirc{}\nabla_k\scirc{}\nabla_\ell
+
\nabla_j\scirc{}\nabla_k\scirc{}P_3^{jk\ell}\scirc{}\nabla_\ell
\right)
\label{Quantization3}
\end{equation}
as a ``minimal'' choice to insure the symmetry of the resulting operator.

\begin{rmk}
{\rm
The formula (\ref{Quantization3}) precisely coincides with the 
projectively equi\-variant quantization \cite{Bou} $\cQ_{0,1}(P):\cF_0(M)\to\cF_1(M)$ 
restricted to cubic polynomials.
}
\end{rmk}

The previously mentioned commutator is actually given by
\begin{eqnarray}
[\widehat{P}_2,\widehat{Q}_2]
&=&
\nonumber
[P_2,Q_2]_S^{jk\ell}\nabla_j\scirc\nabla_k\scirc\nabla_\ell\\[6pt]
&&\label{CommP2Q2}
+\frac{3}{2}\left(\nabla_j[P_2,Q_2]_S^{jk\ell}\right)\nabla_k\scirc\nabla_\ell\\[6pt]
&&
\nonumber
+\left[\frac{1}{2}
\left(\nabla_j\nabla_k[P_2,Q_2]_S^{jk\ell}\right)
+\frac{2}{3}\left(\nabla_k B_{P_2,Q_2}^{k\ell}\right)\right]\nabla_\ell
\end{eqnarray}
where the skew-symmetric tensor
\begin{eqnarray}
B_{P,Q}^{jk}
&=&\nonumber
P^{\ell[j}\nabla_\ell\nabla_m{}Q^{k]m}
+
P^{\ell[j}R^{k]}_{~m,n\ell}Q^{mn}
-(P\leftrightarrow{}Q)\\[6pt]
&&
-\nabla_\ell{}P^{m[j}\nabla_m{}Q^{k]\ell}
-P^{\ell[j}R_{\ell{}m}Q^{k]m}
\label{B}
\end{eqnarray}
satisfies, in addition, $B_{P,Q}=-B_{Q,P}$. We have used the following convention for 
the Riemann and Ricci tensors, viz 
$R^{\ell}_{~i,jk}=\partial_j\Gamma^\ell_{ik} -(j\leftrightarrow{}k) +\ldots$, and 
$R_{ij}=R^{k}_{~i,kj}$.

We can rewrite the commutator (\ref{CommP2Q2}) with the help of the quantization 
prescription (\ref{Quantization012}) and (\ref{Quantization3}) as 
\begin{equation}
[\widehat{P}_2,\widehat{Q}_2]
=
i\widehat{\{P_2,Q_2\}}
+i\widehat{A}_{P_2,Q_2}
\label{CommP2Q2bis}
\end{equation}
where 
\begin{equation}
A_{P_2,Q_2}
=
-\frac{2}{3}\left(\nabla_k B_{P_2,Q_2}^{k\ell}\right)\xi_\ell
\label{A1}
\end{equation}
is a divergence-free vector-field associated with the tensor (\ref{B}) and providing 
the potential quantum correction for quadratic polynomials; recall that, according to~(\ref{Quantization012}), one has $\widehat{A}_{P_2,Q_2}=(i/2)(A_{P_2,Q_2}^\ell\scirc\nabla_\ell+\nabla_\ell\scirc{}A_{P_2,Q_2}^\ell)$.

\goodbreak

We thus have the
\begin{pro}
The commutator of the quantum operators $\widehat{P}$ and $\widehat{Q}$ associated 
with two general quadratic polynomials $P=P_2+P_1+P_0$ and $Q=Q_2+Q_1+Q_0$ reads
\begin{equation}
\frac{1}{i}[\widehat{P},\widehat{Q}]
=
\widehat{\{P,Q\}}
+
\widehat{A}_{P_2,Q_2}
+
\widehat{A}_{P_1,Q_2}-\widehat{A}_{Q_1,P_2}
\label{CommPQ}
\end{equation}
where the third-order differential operator $\widehat{\{P,Q\}}$ is given by 
(\ref{Quantization3}).
\end{pro}
\begin{proof}
The formula (\ref{CommPQ}) results trivially from the previously computed commutators 
and from collecting the anomalous terms appearing in (\ref{CommP1Q2}) and 
(\ref{CommP2Q2bis}) only.
\end{proof}

\begin{rmk}
{\rm
In the special case where $Q_2=H$ as given by (\ref{H}), the anomalous tensor (\ref{B}) 
takes the form
$$
B^{jk}_{P,H}
=
-\half\nabla^{[j}\nabla_\ell P^{k]\ell}
- P^{\ell[j}R^{k]}_\ell
$$
and reduces to 
\begin{equation}
B^{jk}_{P,H}
=
- P^{\ell[j}R^{k]}_\ell
\label{CarterAnomaly}
\end{equation}
if $P$ is a Killing tensor \cite{BCar3}.
}
\end{rmk}

\begin{rmk}
{\rm
In the particular case where $H=\half{}\rg^{jk}(\xi_j-eA_j)(\xi_k-eA_k)$ is the 
Hamiltonian of the electromagnetic coupling, our quantum commutator (\ref{CommPQ}) 
reduces to Carter's formula (6.16) in \cite{BCar3}.
}
\end{rmk}

The purpose of our article is, indeed, to study, using explicit examples, how classical integrability behaves under the ``minimal'' quantization rules proposed in~\cite{BCar3} and somewhat extended here. The next section will be devoted to the computation of the 
quantum corrections in (\ref{CommP1Q2}) and (\ref{CommP2Q2bis}) for all the examples that 
have been previously introduced.

\subsection{The equivariance Lie algebra}\label{EquivCh}

So far, the transformation property of the quantization rules (\ref{Quantization012}) and (\ref{Quantization3}) under a change of coordinates has been put aside. It is mandatory to investigate if these rules are consistent with the map $\cQ:P\mapsto\widehat{P}$ (which has been defined for cubic polynomials, $P=\sum_{k=0}^3{P^{i_1\cdots\i_k}\xi_{i_1}\ldots\xi_{i_k}}$, only) being equivariant with respect to some Lie subgroup of the group of diffeomorphisms of configuration space, $M$. 

Restricting considerations to the infinitesimal version of the sought equivariance, we will therefore look for the set $\fg$ of all vector fields $X$ with respect to which our quantization is equivariant, namely
$L_X\cQ=0$.
From its very definition, $\fg$ is a Lie subalgebra of the Lie algebra, $\Vect(M)$, of vector fields of $M$.
The previous condition means that, for each polynomial $P$, the following holds:
\begin{equation}
L_X(\cQ(P)\phi)-\cQ(L_X{P})\phi-\cQ(P)L_X\phi=0
\label{EquivCondoriginal}
\end{equation}
where $L_X\phi$ denotes the Lie derivative of the half-density $\phi$ of $M$ with respect to the vector field $X\in\fg$ and $L_X{P}=\{X,P\}$ is the Poisson bracket of $X=X^i\xi_i$ and $P$.

Let us recall that, putting locally $\phi=f\vert\vol\vert^\half\in\cF_\half$ with $f\in{}C^\infty(M)$, we get the following expression for the Lie derivative: $L_X\phi=(Xf+\half\rdiv(X)f)\vert\vol\vert^\half$, or with a slight abuse of notation,
$L_X\phi=X^j\nabla_j\phi+\half(\nabla_jX^j)\phi=\half(X^j\scirc\nabla_j+\nabla_j\scirc{}X^j)\phi$, that is
\begin{equation}
L_X\phi=\frac{1}{i}\widehat{X}\phi
\end{equation}
for any $X\in\Vect(M)$.

The equivariance condition (\ref{EquivCondoriginal}) must hold for any $\phi\in\cF_\half$ and thus translates into
\begin{equation}
[\widehat{X},\widehat{P}]=i\widehat{\{X,P\}}
\label{EquivCond}
\end{equation}
for any $X\in\fg$ and any cubic polynomial $P$. The Condition (\ref{EquivCond}) characterizes the Lie algebra $\fg$ we are looking for. We will consider successively the case of polynomials of increasing degree:

(i) Returning to the previous relations (\ref{CommP0Q1}), (\ref{CommP1Q1}) together with $X=P_1$ and $P=Q_0+Q_1$, we readily find that the Lie algebra $\fg_1$ spanned by the solutions of~(\ref{EquivCond}) restricted to polynomials $P$ of degree one is $\fg_1=\Vect(M)$.  

(ii) Let us now proceed to the case of quadratic polynomials $P=P^{jk}\xi_j\xi_k$. The relations (\ref{CommP1Q1}) and (\ref{A0}) give, in that case, the following equivariance defect
\begin{equation}
[\widehat{X},\widehat{P}]-i\widehat{\{X,P\}}
=
\frac{i}{2}\nabla_j \scirc P^{jk}\scirc\nabla_k(\nabla_\ell X^\ell)\bone.
\label{EquivCond2}
\end{equation}
This defect vanishes for any such $P$ iff $\nabla_k(\nabla_\ell X^\ell)=0$, i.e.,
\begin{equation}
d(\rdiv(X))=0.
\label{EquivCond2bis}
\end{equation}
The vector fields $X$ with constant divergence span now a subspace $\fg_2\subset\fg_1$ which is, indeed, an infinite dimensional Lie subalgebra of $\Vect(M)$. The ``minimal'' quantization restricted to quadratic polynomials is therefore equivariant with respect to the group of all diffeomorphisms which preserve the volume up to a multiplicative nonzero constant.

(iii) Let us finally consider homogeneous cubic polynomials $P=P^{jk\ell}\xi_j\xi_k\xi_\ell$ and compute the equivariance defect in this case. A tedious calculation leads to
\begin{equation}
[\widehat{X},\widehat{P}]-i\widehat{\{X,P\}}
=
i\widehat{Z},
\qquad
Z=Z^j\xi_j,
\label{EquivCond3}
\end{equation}
with
\begin{equation}
Z^j
=
\nabla_k\left[
P^{jk\ell}\nabla_\ell\rdiv(X)-P^{\ell{}m[j}L_X\Gamma^{k]}_{\ell{}m}
\right]
\label{Zj}
\end{equation}
where 
\begin{equation}
L_X\Gamma^{k}_{\ell{}m}=\nabla_\ell\nabla_mX^k-R^k_{\ m,n\ell}X^n
\label{LieGamma}
\end{equation}
is the Lie derivative of the symmetric linear connection $\nabla$ with respect to the vector field $X$.
\begin{pro}
The Lie algebra $\fg\subset\Vect(M)$ with respect to which the ``minimal'' quantization (\ref{Quantization012}) and (\ref{Quantization3}) is equivariant is $\mathrm{aff}(M,\nabla)$, the Lie algebra of affine vector fields of $(M,\nabla)$.
\end{pro}
\begin{proof}
The equivariance condition (\ref{EquivCond}), defining the Lie algebra $\fg_3$ we are looking for, is equivalent to $Z=0$ in (\ref{EquivCond3}) for all symmetric tensor fields $P^{jk\ell}$, i.e., thanks to (\ref{Zj}) to
$$
T_k^{jk\ell}\nabla_\ell\rdiv(X)-T_k^{\ell{}m[j}L_X\Gamma^{k]}_{\ell{}m}
=
0
$$
for all tensor fields $T_k^{\ell{}mj}=T_k^{(\ell{}mj)}$. This readily implies that
$$
2\delta^j_{(i}\delta^k_\ell\nabla_{m)}\rdiv(X)
+\delta^j_{(i}L_X\Gamma^{k}_{\ell{}m)}
-\delta^k_{(i}L_X\Gamma^{j}_{\ell{}m)}
=
0.
$$
Summing over $i=j$, one gets 
$$
2n\delta^k_m\nabla_{\ell}\rdiv(X)
+4\delta^k_\ell\nabla_{m}\rdiv(X)
+(n+1)L_X\Gamma^{k}_{\ell{}m}
-\delta^k_{m}L_X\Gamma^{i}_{\ell{}i}
-\delta^k_{\ell}L_X\Gamma^{i}_{m{}i}
=
0,
$$
where $n=\dim(M)$, hence $\nabla_{i}\rdiv(X)=0$ and $L_X\Gamma^{k}_{ij}=\delta^k_i\varphi_j+\delta^k_j\varphi_i$ for some $1$-form~$\varphi$ depending upon the (projective) vector field $X$. The expression (\ref{LieGamma}) of the Lie derivative of the symmetric connection $\nabla$ then yields  $L_X\Gamma^j_{ij}=(n+1)\varphi_i=0$ since we have found that $\nabla_{i}\nabla_j X^j=0$. This entails $L_X\Gamma^{k}_{ij}=0$, proving that $\fg=\fg_3$ is nothing but the Lie algebra $\mathrm{aff}(M,\nabla)$ of affine vector fields.
\end{proof}

We thus obtain the nested equivariance Lie algebras
$$
\fg=\mathrm{aff}(M,\nabla)\subset\fg_2\subset\fg_1=\Vect(M)
$$
where $\fg_2$ is the Lie algebra of vector fields with constant divergence. (Note that if~$M$ is compact without boundary, $\fg_2$ reduces to the Lie algebra of divergence-free vector fields.)

Conspicuously, our quantization scheme turns out to be equivariant with respect to a rather small Lie subgroup of $\Diff(M)$, namely of the affine group of $(M,\nabla)$. It would be interesting to investigate to what extent the equivariance under the sole affine group, $\GL(n,\bR)\ltimes\bR^n$, of a flat affine structure $(M,\nabla)$ allows one to uniquely extend to the whole algebra of polynomials the quantization scheme we have devised for cubic polynomials.

\subsection{The quantum St\"ackel system}\label{QStaeckel}

The quantization of the general St\"ackel system (see Section \ref{StaeckelSection}) 
has first been under\-taken by Benenti, Chanu and Rastelli in \cite{BCR1,BCR2}. We 
will derive, here, the covariant expression of the quantum correction associated to the 
``minimal'' quantization, with the help of the results obtained in 
Section \ref{SectionQuant}.

Denote by $I_i=I_{2,i}+I_{0,i}$ the $i$-th St\"ackel conserved quantity, 
$i=1,\ldots,n$, in~(\ref{Staeckelconserved}) where the indices $0$ and $2$ refer 
to the degree of homogeneity with respect to the coordinates $\xi$. 
Applying (\ref{CommPQ}) with $P_1=Q_1=0$, $P_2=I_{2,i}$ and $Q_2=I_{2,j}$ one gets 
$$
[\widehat{I}_i,\widehat{I}_j]
=
[\widehat{I}_{2,i},\widehat{I}_{2,j}]
=
i\widehat{A}_{I_{2,i},I_{2,j}
}
=
\frac{2}{3}\left(\nabla_kB_{I_{2,i},I_{2,j}}^{k\ell}\right)\nabla_\ell.
$$
\begin{rmk}\label{potnotanomalous}
{\rm This result shows that there are no quantum corrections produced by the potential term. 
More generally, start with a system defined by independent, homogeneous, quadratic observables $H_1,\ldots,H_n$ which is integrable at the classical and quantum levels. Consider a new set of observables $H_1+U_1,\ldots,H_n+U_n$ obtained by adding potential terms $U_1,\ldots,U_n$; if the new system is classically integrable, it will remain integrable at the quantum level.}
\end{rmk}

We are now in position to prove the following
\begin{pro}\label{StaeckelAnomaly}
The quantum correction (\ref{B}) of a general St\"ackel system, with commuting conserved 
quantities $I_1,\ldots,I_n$ defined by (\ref{Staeckelconserved}), retains the form
\begin{equation}
B_{I_{2,i},I_{2,j}}^{k\ell}
=
-2I_{2,i}^{s[k}R_{s{}t}I_{2,j}^{\ell]t}
\label{BStaeckel}
\end{equation}
for $i,j=1,\ldots,n$, where $R_{st}$ denotes the components of the Ricci tensor of 
the metric associated with the Hamiltonian $I_1$.
\end{pro}

\begin{proof}
As a preliminary remark, let us observe that the St\"ackel metric, given by 
(\ref{HStaeckel}), needs not be Riemannian. So we will write it
\begin{equation}
\rg=\sum_{i=1}^n{\frac{(dx^i)^2}{A^i_1(x)}}=\sum_{a=1}^n{\eta_a(\theta^a)^2}
\label{metricStaeckel}
\end{equation}
where $(\theta^a=dx^a/\sqrt{\vert{}A^a_1\vert})_{a=1,\ldots,n}$ is the orthonormal 
moving coframe and the signature of $\rg$ is given by $\eta_a=\mathrm{sign}(A^a_1)$. We will denote by 
$(e_a=\sqrt{\vert{}A^a_1\vert}\partial_a)_{a=1,\ldots,n}$ the associated 
orthonormal frame with respect to the metric $\eta_{ab}=\eta_a\delta_{ab}$ used 
to raise and lower frame indices.

Let us recall, in order to fix the notation, that the connection form $\omega$ 
satisfies the structure equation $d\theta^a+\omega^a_{\ b}\wedge\theta^b=0$ and 
the associated curvature form, $\Omega$, given by 
$\Omega^a_{\ b}=d\omega^a_{\ b}+\omega^a_{\ c}\wedge\omega^c_{\ b}$, is expressed 
in terms of the Riemann tensor by 
$\Omega^a_{\ b}=\half{}R^a_{\ b,cd}\,\theta^c\wedge\theta^d$. The indices $a,\ldots,d$ 
run from $1$ to $n$ and the Einstein summation convention is used when no 
ambiguity arises. Denoting by $R^\ell_{\ i,jk}$ the local components of the Riemann tensor, we have $R^a_{\ b,cd}=\theta^a_\ell\,R^\ell_{\ i,jk}\,e_b^ie_c^je_d^k$.

We start off with the calculation of the connection form, $\omega$, and of some 
components of the curvature form, $\Omega$. Straightforward computation, using 
relation (\ref{idStaeckel}), then yields 
for the non-vanishing components of the connection
$$
\omega_{ab,a}=\half\eta_b\,C^a_b\frac{\left|A^b_1\right|^{3/2}}{\left|A^a_1\right|},
\quad a\neq b,\qquad 
\omega_{ab,c}=\omega_{ab}(e_c), 
$$
the other nontrivial components $\omega_{ab,b}$ are obtained accordingly.
For the curvature, a lengthy computation gives the special components
\begin{equation}\label{courbure}
R_{ac,cb}
=
3\left(
-\eta_a\omega_{ca,c}\,\omega_{ab,a}
-\eta_b\omega_{cb,c}\,\omega_{ba,b}
+\eta_c\omega_{ca,c}\,\omega_{cb,c}
\right),\qquad a\neq b,
\end{equation}
which will be needed in the sequel.

Two last ingredients are the introduction of the frame components of various objects. 
We will denote the Killing tensor $I_{2,i}$ (resp. $I_{2,j}$) as $P$ (resp. $Q$). 
Their frame components $P=P^{bc}\,e_b\otimes{}e_c$, and similarly for $Q$, will be
\begin{equation}\label{frameKilling}
P^{bc}=p_b \d_{bc},\qquad p_b=\frac{A_{i}^b}{2\vert{}A_1^b\vert},\qquad 
Q^{bc}=q_b \d_{bc},\qquad q_b=\frac{A_{j}^b}{2\vert{}A_1^b\vert}.
\end{equation}
The covariant derivative will have the frame components 
\[
{\cal D}_c P_{ab}=e_c(P_{ab})-\om^s_{~a,c}P_{sb}-\om^s_{~b,c}P_{as}.
\]
The equations which express that $P^{ab}$ is a Killing tensor are now
\begin{equation}\label{frameKillingeq}
\begin{array}{rcll}
e_b(p_a)&=&2\om_{ab,a}(\eta_a p_a-\eta_b p_b),&a\neq{}b,\\[6pt]
e_a(p_a)&=&0,&
\end{array}
\end{equation}
where the repeated indices are not summed over. One can check that they hold true 
using the explicit form of $p_a$ given in (\ref{frameKilling}) and the 
identity (\ref{idStaeckel}).

Using all of the previous information one can compute the frame components of the 
various pieces appearing in the tensor $B^{ij}_{P,Q}.$ We have successively
\[
\begin{array}{l}
P^{s[i}\nabla_s\nabla_t Q^{j]t}- (P\ \leftrightarrow\ Q)\ =\\[4mm]
\displaystyle\sum_{l\neq{}i,j}(4\om_{li,l}\om_{lj,l}-3\eta_l\eta_i\om_{li,l}\om_{ij,i}
-3\eta_l\eta_j\om_{lj,l}\om_{ji,j})
\left[\rule{0mm}{4mm}p_iq_j-\eta_lp_l\eta_iq_j+\eta_lq_l\eta_ip_j
-(i\leftrightarrow j)\right]
\end{array}
\]
and
\[
\nabla_s P^{t[i}\nabla_t Q^{j]s}=\half\sum_l\om_{li,l}\om_{lj,l}
\left[\rule{0mm}{4mm}p_iq_j-\eta_lp_l\eta_iq_j+\eta_lq_l\eta_ip_j
-(i\ \leftrightarrow\ j)\right].
\]
Combining these relations, and using (\ref{courbure}), we get
\[
\begin{array}{l}
P^{s[i}\nabla_s\nabla_t Q^{j]t}- (P\ \leftrightarrow\ Q)
-\nabla_s P^{t[i}\nabla_t Q^{j]s}\ =\\[4mm]
\displaystyle
\hspace{4cm} \half\sum_l\eta_l R_{il,lj}
\left[\rule{0mm}{4mm}p_iq_j-\eta_lp_l\eta_iq_j+\eta_lq_l\eta_ip_j
-(i\ \leftrightarrow\ j)\right].
\end{array}
\]
Let us then compute
\[
P^{s[i}R^{j]}_{~u,vs}Q^{uv}-(P\ \leftrightarrow\ Q)
=
\half\sum_l\eta_l R_{il,lj}
\left[\rule{0mm}{4mm}\eta_lp_l\eta_iq_j-\eta_lq_l\eta_ip_j
-(i\ \leftrightarrow\ j)\right].
\]
Collecting all the pieces leaves us with
\begin{equation}\label{anomalie}
\begin{array}{l}
P^{s[i}\nabla_s\nabla_t Q^{j]t}+P^{s[i}R^{j]}_{~u,vs}Q^{uv}- (P\ \leftrightarrow\ Q)
-\nabla_s P^{t[i}\nabla_t Q^{j]s}\ =\\[4mm]
\hspace{4cm}\half\sum_l\eta_lR_{il,lj}(p_iq_j-p_jq_i).
\end{array}
\end{equation}
The last sum is nothing but the frame components of the tensor 
$-P^{s[i}R_{st}Q^{j]t},$ so that we have obtained the tensorial relation
\begin{equation}\label{cunu4}
P^{s[i}\nabla_s\nabla_t Q^{j]t}+P^{s[i}R^{j]}_{~u,vs}Q^{uv}- (P\ \leftrightarrow\ Q)
-\nabla_s P^{t[i}\nabla_t Q^{j]s}=-P^{s[i}R_{st}Q^{j]t},
\end{equation}
which implies
\begin{equation}
B^{ij}_{P,Q}=-2P^{s[i}R_{st}Q^{j]t},
\label{Anomaly22}
\end{equation}
in agreement with  \cite{BCR2}. This ends the proof of Proposition \ref{StaeckelAnomaly}.
\end{proof}

Now we can come to the central point of our analysis: is a St\" ackel system integrable 
at the quantum level? The answer is given by the following 
\begin{cor}(\cite{BCR1,BCR2})
A St\" ackel system is integrable at the quantum level iff
\begin{equation}
R_{ij}=0\qquad\mbox{for}\ i\neq j,\ \mathrm{where}\ i,j=1,\ldots,n,
\label{RicDiag}
\end{equation}
in the special coordinates which are constituent to this system.
\end{cor}
\begin{proof}
The Killing tensors $I_{2,i}$ are diagonal, for $i=1,\ldots,n$, in the St\"ackel coordinate system, and the proof follows from (\ref{BStaeckel}). 
\end{proof}
The conditions (\ref{RicDiag}) are known as the Robertson conditions \cite{HPRob}, as interpreted by Eisenhart~\cite{LPEis}. Quite recently, Benenti et al \cite{BCR1} have refined the definition of the separability of the Schr\"odinger equation and shown that, for St\"ackel systems, the Robertson conditions are necessary and sufficient for the 
separability of the Schr\"odinger equation. As mentioned in Remark \ref{RmkPars}, the classical integrability is equivalent to the separability of the Hamilton-Jacobi equation; the situation for these systems can be therefore summarized by the following diagram:

\goodbreak

\[\mbox{Classical integrability}\quad\Longleftrightarrow
\quad\mbox{separable Hamilton-Jacobi }\]

\[\Downarrow\qquad \mbox{provided}\qquad R_{ij}=0\quad(i\neq j)\]

\[\mbox{Quantum integrability}\quad\Longleftrightarrow
\quad\mbox{separable Schr\"odinger}\]

\subsection{The quantum ellipsoid and Neumann systems}
\label{QEllipsoidNeumann}

It is now easy to prove that the ellipsoid geodesic flow 
(see section \ref{subellipsoid}), including the potential given in 
(\ref{HamiltonianJacobi}), is integrable at the quantum level. Using the 
coordinates~$(x^i)$ and the (Riemannian) metric given by 
(\ref{MetricEllipsoid}), one can check that the 
Ricci tensor has components 
\[
R_{ij}=\frac{\cal N}{x^i} \sum_{s\neq i}\frac 1{x^s}\,\rg_{ij},\qquad \quad
{\cal N}=\frac{a_0a_1\cdots a_n}{x^1\cdots x^n},
\]
and therefore satisfies the Robertson conditions. As already emphasized, the occurrence 
of an additional potential is irrelevant for the quantum analysis since the potential terms do not generate quantum corrections (see Remark \ref{potnotanomalous}).

Similarly we get the quantum integrability for the Neumann system (see Section~\ref{subNeumann}) using the metric on $S^n$ given by (\ref{metricNeumann}). The Ricci tensor being given by
\[
R_{ij}=(n-1)\rg_{ij},
\]
the Robertson conditions are again satisfied.

\subsection{The quantum generalized Kerr-Newman system}\label{QKNS} 

The quantization of the four commuting observables (\ref{HP-KNS}) and (\ref{ST-KNS}) is 
straightforward.

In view of the relations given in Section \ref{QuantIntSys} all quantum commutators 
vanish except for $[\widehat{\widetilde{H}},\widehat{\widetilde{P}}]$; this is due to 
the fact that the conserved quantities $\widetilde{S}$ and $\widetilde{T}$ (see~(\ref{ST-KNS})) are Killing-Maxwell vector fields.

The anomalous terms in the previous commutator are $A_{P_2,H_2}$, 
$A_{P_1,H_2}$ and $A_{P_2,H_1}$ where $P_2=P^{ij}\xi_i\xi_j$, 
$H_2=\half\rg^{ij}\xi_i\xi_j$, $P_1=-2P^{ij}\xi_i A_j$ and $H_1=-\rg^{ij}\xi_i A_j$. 

The vector field $A_{P_2,H_2}$ given by (\ref{A1}) actually vanishes because, cf. 
(\ref{CarterAnomaly}), $B^{jk}_{P_2,H_2}
=
- P^{\ell[j}R^{k]}_\ell=0$ as a consequence of (\ref{KillingMaxEqs}); indeed the tensor $P$ anti-commutes with the electromagnetic field strength $F$, implying that it commutes with the stress-energy electromagnetic tensor, hence with the Ricci tensor in view of the Einstein-Maxwell equations \cite{BCar3}.

The two other anomalous terms (\ref{A0}) also vanish as it turns out that 
$\nabla_jA^j=0$ (see~(\ref{DivA})) and $\nabla_j(P^{jk}A_k)=0$.

This derivation reproduces and extends Carter's results to the generalized Kerr-Newman solution, in a somewhat shorter manner.

\begin{rmk}
{\rm
Our analysis of quantum integrability for the generalized Kerr-Newman solution in $4$ dimensions can be carried over into recent work \cite{GHY,KL,SY} dealing with $5$-dimensional black holes. In these cases, classical integrability follows from the existence of $3$ Killing vectors and $1$ quadratic Killing tensor, besides the Hamiltonian. These metrics being Einstein, the above arguments given for the generalized Kerr-Newman case apply just as well, insuring quantum integrability. This fact is in agreement with the separability of the Laplace operator.
}
\end{rmk}

\goodbreak

\subsection{The quantum Multi-Centre system}\label{QMC} 
For this example too, the quantization is straightforward. The single point to be
checked for quantum integrability is just the commutator $[\widehat{H},\widehat{P}],$ 
with the possible quantum correction~(\ref{CarterAnomaly}) given by $- P^{\ell[j}R^{k]}_\ell.$ 
Here it vanishes trivially since these metrics are Ricci-flat.

\goodbreak

\subsection{The quantum Di Pirro system}\label{QuantDiPirro}

As seen in Section \ref{ClassicalDiPirro}, the classical integrability of this system 
is provided by three commuting observables: on the one hand $H$, $P$ respectively 
given by (\ref{HDiPirro}) and (\ref{PDiPirro}), and $T=\xi_3$ if $c(x^3)=\const.$, 
and on the other hand $H$, $P$ and $J=\xi_1{}x^2-\xi_2{}x^1$ if $a=b,\g$ depend 
on $r$ only.

At the quantum level, the Killing vectors $\widehat{T}$ and $\widehat{J}$ do 
commute with $\widehat{H}$ according to (\ref{CommP1Q2}) and (\ref{A0}). As for 
the commutator $[\widehat{P},\widehat{H}]$ of the quantized Killing tensors, it 
is given by  (\ref{CarterAnomaly}), namely $B_{P,H}
=
-\frac{1}{2}P^{\ell[j}R^{k]}_\ell\,\partial_j\wedge\partial_k$, and one finds 
\begin{eqnarray*}
B_{P,H}
&=&
-\frac{3}{16}\,\frac{c'(x^3)}{(\gamma(x^1,x^2)+c(x^3))^3}
(
a(x^1,x^2)\partial_1\g(x^1,x^2)\,\partial_1\wedge\partial_3\\[6pt]
&&+\;b(x^1,x^2)\partial_2\g(x^1,x^2)\,\partial_2\wedge\partial_3
).
\end{eqnarray*}

\goodbreak

For the system $(H,P,T)$, this quantum correction vanishes since $c'(x^3)=0$, implying quantum integrability. However, for the system $(H,P,J)$, in the generic case $\g\neq\const.$, 
we get $B_{P,H}\neq0$, showing that the minimal 
quantization rules may produce quantum corrections.

\goodbreak

\section{Discussion and outlook}\label{ChConclusion}

It would be worthwhile to get insight into the status of our ``minimal'' quantization 
rules and to their relationship with other bona fide quantization procedures. Among 
the latter, let us mention those obtained by geometric means, and more 
specifically by imposing equivariance of the quantization mapping, $\cQ$, with 
respect to some symmetry group, $G$, e.g., a group of automorphisms of a certain 
geometric structure on configuration space, $M$. We refer to the articles 
\cite{LO,DO1,DO2,DLO,Bor} for a detailed account on equivariant quantization. 
The two main examples are respectively the projectively, $G=\mathrm{SL}(n+1,\bR)$, 
and conformally, $G=\mathrm{O}(p+1,q+1)$, equivariant quantizations which have 
been shown to be uniquely determined \cite{LO,DLO,DO1,DO2}. For instance, the conformally 
equivariant quantization $\cQ_{\half}:\cF_\half(M)\to\cF_\half(M)$ has 
been explicitly computed for quadratic \cite{DO1} and cubic \cite{LD} 
observables; for example, if $P=P^{ij}\xi_i\xi_j$ we then have
\begin{equation}
\cQ_{\half}(P)
=
\widehat{P}
+
\beta_3\,\nabla_i\nabla_j(P^{ij})+\beta_4\,\rg^{ij}\rg_{k\ell}\nabla_i\nabla_j(P^{k\ell})
+\beta_5\,R_{ij}P^{ij}+\beta_6\,R\rg_{ij}P^{ij}
\label{Qhalf}
\end{equation}
where the ``minimal'' quantum operator
\begin{equation}
\widehat{P}=-\nabla_i\scirc{}P^{ij}\scirc\nabla_j
\label{ourPhat}
\end{equation}
is given by (\ref{Quantization012}), together with $\beta_3=-n/(4(n+1))$, 
$\beta_4=-n/(4(n+1)(n+2))$, $\beta_5=n^2/(4(n-2)(n+1))$, 
$\beta_6=-n^2/(2(n^2-4)(n^2-1))$, assuming $n=\dim(M)>2$. In (\ref{Qhalf}) 
we denote by $R_{ij}$ the components of the Ricci tensor and  by $R$ the scalar
curvature. The formula (\ref{Qhalf}) provides a justification of the term ``minimal'' 
for the mapping $P\mapsto\widehat{P}$ given by (\ref{Quantization012}) 
and (\ref{Quantization3}).

We have checked that, in the special instance of the geodesic flow of the 
ellipsoid discussed in Section \ref{subellipsoid}, the quantum commutators 
of the observables $I_i$ defined in~(\ref{PiEllipsoid}), namely 
$[\cQ_{\half}(I_i),\cQ_{\half}(I_j)]$, fail to vanish for 
$i\neq{}j=1,\ldots,n$. Had we started from the expression (\ref{Qhalf}) with 
adjustable coefficients $\beta_3,\ldots,\beta_6$, the requirement that the 
latter commutator be vanishing imposes $\beta_3=\ldots=\beta_6=0$, leading 
us back to the minimal quantization rule (\ref{ourPhat}).

Despite their nice property of preserving, to a large extent, integrability 
(from classical to quantum),  the ``minimal'' quantization rules still remain 
an ad hoc procedure, defined for observables at most cubic in 
momenta, and do not follow from any sound constructive principle, be it of 
a geometric or an algebraic nature. The quest for a construct leading unambiguously to a genuine ``minimal'' quantization procedure remains an interesting challenge. As discussed in Section \ref{EquivCh}, the equivariance assumption with respect to the affine group might be helpful for determining the sought ``minimal'' quantization of polynomials of higher degree. This analysis is required for the quantization of, e.g., the newly discovered integrable systems \cite{DM} which involve cubic Killing tensors.

Another field of applications of the present work could be the search for quantum integrability of the geodesic flow on the higher dimensional generalizations of the Kerr metric which have been lately under intense study \cite{GHY,CGLP,VSP}.

Still another perspective for future work would be to generalise the previous computation 
of quantum corrections to the case of classical integrability in the presence of an 
electromagnetic field in a purely gauge invariant manner. In particular the 
approach presented in Section \ref{MaxwellKilling} should be further extended 
at the quantum level via the quantization of the Schouten-Maxwell brackets.



\begin{thebibliography}{99}

\bibitem{BBT}
O. Babelon, D. Bernard, M. Talon,
{\em Introduction to Classical Integrable Systems},
Cambridge University Press (2003).


\bibitem{BT}
M. Bellon, M. Talon,
{\it Spectrum of the quantum Neumann problem},
\texttt{arXiv:hep-th/0407005}.


\bibitem{Ben}
S. Benenti,
{\it Inertia tensors and St\"ackel systems in the Euclidean spaces}, Rend. Sem. 
Mat. Univ. Pol. Torino {\bf 50} (1992) 315--341.

\bibitem{BCR1}
S. Benenti, C. Chanu, and G. Rastelli,
{\it Remarks on the connection between the additive separation of the Hamilton-Jacobi 
equation and the multiplicative separation of the Schr\"odinger equation. I. The 
completeness and Robertson conditions}, J. Math. Phys. {\bf 43} (2002) 5183--5222.

\bibitem{BCR2}
S. Benenti, C. Chanu, and G. Rastelli,
{\it Remarks on the connection between the additive separation of the Hamilton-Jacobi 
equation and the multiplicative separation of the Schr\"odinger equation. II. First 
integrals and symmetry operators}, J. Math. Phys. {\bf 43} (2002) 5223--5253.

\bibitem{Bor}
M. Bordemann,
{\it Sur l'existence d'une prescription d'ordre naturelle projectivement invariante},
\texttt{math.DG/0208171}.

\bibitem{Bou}
S. Bouarroudj, {\it Projectively equivariant quantization map},
Lett. Math. Phys. {\bf 51}:4 (2000) 265--274.

\bibitem{MCar}
M. Cariglia,
{\it Quantum Mechanics of Yano tensors: Dirac equation in curved spacetime}, 
Class. Quant. Grav. {\bf 21} (2004) 1051--1078.

\bibitem{BCar1}
B. Carter,
{\it Hamilton-Jacobi and Schr\"odinger separable solutions of Einstein's equations},  Comm. Math. Phys.  {\bf 10}  (1968) 280--310. 

\bibitem{BCar2}
B. Carter,
{\it Black hole equilibrium states} in Black holes/Les astres occlus (Ecole d'\'et\'e de Physique Th\'eorique, Les Houches, 1972),  pp. 57--214. Gordon and Breach, New York (1973).
 	
\bibitem{BCar3}
B. Carter,
{\it Killing tensor quantum numbers and conserved currents in curved space}, 
Phys. Rev. D{\bf 16} (1977) 3395--3414.

\bibitem{CGLP}
Z. W. Chong, G. W. Gibbons, H. L\"u, C. N. Pope,
{\it Separability and Killing Tensors in Kerr-Taub-NUT-de Sitter Metrics in Higher Dimensions}, 
\texttt{arXiv:hep-th/0405061}.

\bibitem{CFH}
B. Cordani, L. Feh\'er, P. Horv\'athy,
{\it $\mathrm{O}(4,2)$ dynamical symmetry of the Kaluza-Klein monopole}, 
Phys. Lett. B{\bf 201} (1988) 481--486.

\bibitem{DM}
H. R. Dullin, V. S. Matveev,
{\it A new integrable system on the sphere},
Math. Res. Lett.  {\bf 11}  (2004) 715--722. 

\bibitem{DO1}
C. Duval, V. Ovsienko,
{\it Conformally equivariant quantum Hamiltonians},
Selecta Math. (N.S.) {\bf 7}:3 (2001) 291--320.

\bibitem{DO2}
C. Duval, V. Ovsienko,
{\it Projectively equivariant quantization and symbol calculus:
noncommutative hypergeometric functions}, Lett. Math. Phys. {\bf 57}:1 (2001)
61--67.

\bibitem{DLO}
C. Duval, P. Lecomte, V. Ovsienko,
{\it Conformally equivariant quantization: existence and uniqueness},
Ann. Inst. Fourier. {\bf  49}:6 (1999) 1999--2029.

\bibitem{LPEis}
L. P. Eisenhart, 
{\it Separable systems of St\"ackel}, 
Ann. of Math. {\bf 35}:2 (1934) 284--305.

\bibitem{GHY}
G. W. Gibbons, S. A. Hartnoll, Y. Yasui,
{\it Properties of some five-dimensional Einstein metrics},
Class. Quant. Grav. {\bf 21} (2004) 4697--4730.

\bibitem{GR}
G. W. Gibbons, P.J. Ruback,
{\it The hidden symmetries of Multi-Centre metrics},
Comm. Math. Phys. {\bf 115} (1988) 267--300.

\bibitem{GRvH}
G. W. Gibbons, R. H. Rietdijk, J. W. van Holten,
{\it SUSY in the sky},
Nucl. Phys. {\bf 404} (1993) 42.

\bibitem{HW}
J. Harnad, P. Winternitz,
{\it Classical and quantum integrable systems in $\widetilde{\mathfrak{gl}}(2)^{+*}$ and separation of variables},
Comm. Math. Phys. {\bf 172}:2 (1995) 263--285.

\bibitem{KL}
H. K. Kunduri, J. Lucietti,
{\it Integrability and the Kerr-(A)dS black hole in five dimensions},
\texttt{arXiv:hep-th/0502124}.


\bibitem{LO}
P.B.A. Lecomte and V. Ovsienko,
{\it Projectively invariant symbol calculus}, Lett. Math. Phys. 
{\bf 49}:3 (1999) 173--196.

\bibitem{LD}
S.E. Loubon Djounga, 
{\it Conformally invariant quantization at order three},
Lett. Math. Phys. {\bf 64}:3 (2003) 203--212.

\bibitem{Mos1}
J. Moser,
{\it Various aspects of the integrable Hamiltonian systems}, in Dynamical 
systems (C.I.M.E. Summer School Bressanone; 1978), Progress in Mathematics, 
Birkh\"auser {\bf 8} (1981) 233--289.

\bibitem{Mos2}
J. Moser,
{\it Geometry of quadrics and spectral theory}, in Proceedings of the Chern Symposium, 
Berkeley 1979, Springer (1980) 147--188.


\bibitem{Per}
A.M. Perelomov,
{\em Integrable Systems of Classical Mechanics and Lie Algebras}, Vol I,
Birkh\"auser (1990), and references therein.

\bibitem{Ple}
J. F. Plebanski,
{\it A class of solutions of Einstein-Maxwell equations}, 
Ann. of Phys. {\bf 90} (1975) 196--255.


\bibitem{PD}
J. F. Plebanski and M. Demianski,
{\it Rotating, charged, and uniformly accelerating mass in General Relativity}, 
Ann. of Phys. {\bf 98} (1976) 98--127.

\bibitem{HPRob}
H. P. Robertson, 
{\it Bermerkung \"uber separierbare Systeme in der Wellenmechanik},
Math. Annal. {\bf 98} (1927) 749--752.

\bibitem{SY}
M. Sakaguchi, Y. Yasui,
{\it Notes on Five-dimensional Kerr Black Holes},
\texttt{arXiv:hep-th/0502182}.

\bibitem{Sou}
J.-M.~Souriau,
{\em Structure des syst\`emes dynamiques},
Dunod (1970, \copyright 1969);
{\em Structure of Dynamical Systems. A Symplectic View of Physics},
translated by C.H.~Cushman-de Vries (R.H.~Cushman and G.M.~Tuynman,
Translation Editors), Birkh\"auser (1997).

\bibitem{Tot}
J.A. Toth, 
{\it Various quantum mechanical aspects of quadratic forms},
J. Funct. Anal. {\bf 130}:1 (1995) 1--42.

\bibitem{Val}
G.Valent,
{\it Integrability versus separability for the Multi-Centre metrics},
Comm. Math. Phys. {\bf 244} (2004) 571--594.

\bibitem{VSP}
M. Vasudevan, K. A. Stevens, D. N. Page,
{\it Separability of the Hamilton-Jacobi and Klein-Gordon Equations in Kerr-de Sitter Metrics},
Class.Quant.Grav. {\bf 22} (2005) 339-352.

\bibitem{Wei}
S. Weigert, 
{\it The problem of quantum integrability},
Physica D  {\bf 56}:1  (1992) 107--119.

\end{thebibliography}
\end{document}